\documentclass[12pt,a4paper]{article}
\usepackage{t1enc}
\usepackage[latin1]{inputenc}
\usepackage[english]{babel}
\begin{document}

\baselineskip 0.85cm
\topmargin -0.55in
\oddsidemargin 0in

\let\ni=\noindent

\renewcommand{\thefootnote}{\fnsymbol{footnote}}

\newcommand{ \mi }{ {\stackrel{o}{m}}}

\newcommand{\SM}{Standard Model }

\newcommand{\CKM}{Cabibbo---Kobayashi---Maskawa }

\newcommand{\UK}{Super--Kamiokande }

\newcommand{\mnu }{\nu_s^{(\mu)} }

\newcommand{\enu }{\nu_s^{(e)} }

\pagestyle {plain}

\setcounter{page}{1}

\pagestyle{empty}

~~~
\ \hfill IFT-01/25

\vspace{1cm}

{\large\centerline{\bf An effective model for quark masses and mixings{\footnote{ Work supported in part by the Polish Committee for Scientific Research (KBN), Grant 5 P03B 119 20 (2001 -- 2002).}}}}

\vspace{0.5cm}

{\centerline {\sc Wojciech Kr\'{o}likowski}}

\vspace{0.5cm}

{\centerline {\it Institute of Theoretical Physics, Warsaw University}}

{\centerline {\it Ho\.{z}a 69,~~PL--00--681 Warszawa, ~Poland}}

\vspace{0.5cm}

{\centerline {\bf Abstract}}

\vspace{0.3cm}

By analogy with an effective model of charged-lepton mass matrix that, with the inputs of $m^{\rm exp}_e $ and  $m^{\rm exp}_\mu $, predicts (in a perturbative zero order) $m_\tau = 1776.80 $ MeV close to $m^{\rm exp}_\tau = 1777.03^{+0.30}_{-0.26}$ MeV, we construct such a model for quark mass matrices reproducing consistently the bulk of experimental information on quark masses and mixings. In particular, the model predicts $|V_{u b}| = 0.00313$, $\gamma = - \arg V_{u b} = 63.8^\circ $ and $|V_{t d}| = 0.00785$, $\beta = - \arg V_{t d} = 20.7^\circ $ ({\it i.e.}, $\sin 2\beta = 0.661$ to be compared with the BaBar value $\sin 2\beta^{\rm exp} = 0.59 \pm 0.14$), if the figures $|V^{\rm exp}_{u s}| = 0.2196$, $|V^{\rm exp}_{c b}| = 0.0402$ and $m^{\rm exp}_{s} = 123$ MeV, $m^{\rm exp}_{c} = 1.25$ GeV, $m^{\rm exp}_{b} = 4.2$ GeV are used as inputs.  Also the rest of CKM matrix elements is predicted consistently by the experimental data. Here, quark masses and CKM matrix elements (ten independent quantities) are parametrized by eight independent model constants, what gives two independent predictions, {\it e.g.} for $|V_{ub}|$ and $\beta$. The considered model deals with the fundamental-fermion Dirac mass matrices, so that the neutrino Majorana mass matrix is outside the scheme. Some foundations of the model are collected in Appendix.

\vspace{0.2cm} 

\ni PACS numbers: 12.15.Ff , 14.60.Pq , 12.15.Hh 
 
\vspace{0.5cm} 

\ni August 2001

\vfill\eject

\pagestyle {plain}

\setcounter{page}{1}

~~~

\vspace{0.1cm}

\ni {\centerline{\bf 1. Introduction}}

\vspace{0.3cm}

The explicit effective form of mass matrix invented for three generations of charged leptons $ e^-\,,\,\mu^-\,$, $\tau^- $, and being surprisingly good for their masses [1], is applied in this paper to three generations of up and down quarks, $u\,,\,c\,,\,t $ and $ d\,,\,s\,,\,b $, in order to correlate tentatively their masses and mixing parameters. This form reads

\begin{equation}
\left({M}^{(f)}_{\alpha \beta}\right) = \frac{1}{29} \left(\begin{array}{ccc} \mu^{(f)}\varepsilon^{(f)} & 2\alpha^{(f)} e^{i\varphi^{(f)}} & 0 \\ & & \\ 2\alpha^{(f)} e^{-i\varphi^{(f)}} & 4\mu^{(f)}(80 + \varepsilon^{(f)})/9 & 8\sqrt{3}\,\alpha^{(f)} e^{i\varphi^{(f)}} \\ & & \\ 0 & 8\sqrt{3}\,\alpha^{(f)} e^{-i\varphi^{(f)}} & 24\mu^{(f)} (624 + \varepsilon^{(f)})/25 \end{array}\right) \;,
\end{equation}

\ni where the label $f = e $ or $ u\,,\,d $ is used to denote charged leptons or up and down quarks, respectively, while $\mu^{(f)}$, $\varepsilon^{(f)}$, $\alpha^{(f)}$ and $\varphi^{(f)}$ are real constants to be determined from the present and future experimental data for charged-lepton or quark masses and mixing parameters ($\mu^{(f)}$ and $\alpha^{(f)}$ are mass--dimensional). 

 Here, the form (1) of mass matrix $\left({M}^{(f)}_{\alpha \beta}\right)$ may be considered as a detailed ansatz to be compared with the charged-lepton or quark data. However, in the past, we have presented an argument [2,1] in favour of the form (1), based on: ({\it i}) K\"{a}hler--like generalized Dirac equations (interacting with the Standard Model gauge bosons) whose {\it a priori} infinite sequence is necessarily reduced (in the case of fermions) to three Dirac equations, due to an intrinsic Pauli principle, and ({\it ii}) an ansatz for the Dirac mass matrix, suggested by the above three--generation characteristics ({\it i}). For the reader's convenience this argument is reproduced in Appendix.

 In the case of charged leptons, assuming that the off--diagonal elements of the mass matrix $\left({M}^{(e)}_{\alpha \beta}\right)$ can be treated as a small perturbation of its diagonal terms ({\it i.e.}, that $\alpha^{(e)}/\mu^{(e)}$ is small enough), we calculate in the lowest perturbative order [1]

\vspace{-0.2cm}

\begin{eqnarray}
m_\tau & = & \left[ 1776.80 + 10.2112 \left( \frac{\alpha^{(e)} }{\mu^{(e)} } \right)^2\,\right]\;{\rm MeV} \;, \nonumber \\ 
\mu^{(e)} & = & 85.9924 \; {\rm MeV} + O\left[ \left( \frac{\alpha^{(e)} }{\mu^{(e)} } \right)^2\,\right]\, \mu^{(e)} \;, \nonumber \\ 
\varepsilon^{(e)} & = & 0.172329 + O\left[ \left( \frac{ \alpha^{(e)}}{ \mu^{(e)}}\right)^2 \right]\;,
\end{eqnarray}

\ni when the experimental values of $ m_e $ and $ m_\mu $ [3] are used as inputs. In Eqs. (2), the unperturbed parts are given as $\stackrel{\circ}{m}_\tau = 6(351m_\mu - 136 m_e)/125 $, $\stackrel{\circ}{\mu}^{(e)} = 29(9m_\mu - 4 m_e )/320 $ and $ \stackrel{\circ}{\varepsilon}^{(e)} = 320m_e/(9m_\mu - 4 m_e)$, respectively. We can see that the predicted value of $ m_\tau $ agrees very well with its experimental figure $ m_\tau^{\rm exp} = 1777.03^{+0.30}_{-0.26} $~MeV [3], even in the zero perturbative order. To estimate $\left(\alpha^{(e)}/\mu^{(e)}\right)^2 $, we can take this experimental figure as another input, obtaining

\begin{equation}
\left(\frac{\alpha^{(e)}}{\mu^{(e)}}\right)^2 = 0.023^{+0.029}_{-0.025} \;,
\end{equation}

\ni  which value is not inconsistent with zero. Hence, $\alpha^{(e)\,2} = 170^{+220}_{-190}\, {\rm MeV}^2 $ due to Eq. (2).

For the unitary matrix $\left({U}^{(e)}_{\alpha \beta}\right)$, diagonalizing the charged--lepton mass matrix $\left({M}^{(e)}_{\alpha \beta}\right)$ according to the relation $ U^{(e)\,\dagger}\, M^{(e)}\,U^{(e)} = {\rm diag}(m_e\,,\,m_\mu\,,\,m_\tau)$, we get in the lowest perturbative order 

\vspace{-0.2cm}

\begin{eqnarray}
\lefteqn{\left( U^{(e)}_{\alpha \beta}\right) = } \nonumber \\ & & \left(\!\!\begin{array}{ccc} 1 - \frac{2}{29^2}\left( \frac{\alpha^{(e)} }{m_\mu} \right)^2 & \frac{2}{29} \frac{\alpha^{(e)}}{m_\mu} e^{i\varphi^{(e)}} & \frac{16\sqrt{3}}{29^2} \left(\frac{\alpha^{(e)}}{m_\tau} \right)^2 e^{2i \varphi^{(e)}} \\ & & \\ -\frac{2}{29}\frac{\alpha^{(e)}}{m_\mu} e^{-i\varphi^{(e)}} & 1\! - \!\frac{2}{29^2} \left( \frac{\alpha^{(e)}}{m_\mu}\right)^2\! - \!\frac{96}{29^2}\left(\frac{\alpha^{(e)}}{m_\tau}\right)^2 & \frac{8\sqrt{3}}{29} \frac{\alpha^{(e)}}{m_\tau}e^{i\varphi^{(e)}} \\  & & \\ \frac{16\sqrt{3}}{29^2} \frac{\alpha^{(e)\,2}}{m_\mu\, m_\tau}\,e^{-2i \varphi^{(e)}} & - \frac{8\sqrt{3}}{29} \frac{\alpha^{(e)}}{m_\tau} e^{-i \varphi^{(e)}} & 1 - \frac{96}{29^2} \left( frac{\alpha^{(e)}}{m_\tau} \right)^2 \end{array} \!\!\right).
\end{eqnarray}

\vspace{0.3cm}

\ni {\centerline{\bf 2. Quark mass matrices}}

\vspace{0.3cm}

Now, we will try to apply to quarks the form of mass matrix which was worked out above for leptons. To this end, we conjecture for three generations of up quarks $ u\,,\,c\,,\,t $ and down quarks $ d\,,\,s\,,\,b $ the mass matrices $\left(M_{\alpha \beta}^{(u)}\right)$ and $\left(M_{\alpha \beta}^{(d)}\right) $, respectively, essentially of the form (1), where the label $ f = u\,,\,d $ denotes  up and down quarks. The only modification introduced is a new real constant $ C^{(f)}$ added to $\varepsilon^{(f)}$ in the mass-matrix element $ M^{(f)}_{33}$ which now becomes

\begin{equation}
M^{(f)}_{33} = \frac{24 \mu^{(f)}}{25\cdot 29}\left(624 + \varepsilon^{(f)} + C^{(f)}\right)\;.
\end{equation}

\ni Note that our approach refers to the fermion Dirac mass matrices, leaving the neutrino Majorana mass matrix [4] outside the scheme.

Since for quarks the mass scales $\mu^{(u)}$ and $\mu^{(d)}$ are expected to be even more important than the scale $\mu^{(e)}$ for charged leptons, we assume that the off--diagonal elements of mass matrices $\left(M_{\alpha \beta}^{(u)} \right)$ and $\left(M_{\alpha \beta}^{(d)}\right)$ can be considered as a small perturbation of their diagonal terms. Then, in the lowest perturbative order, we calculate the following mass formulae:

\begin{eqnarray}
m_{u,d} & = & \frac{\mu^{(u,d)} }{29} \varepsilon^{(u,d)} - A^{(u,d)} \left( \frac{\alpha^{(u,d)}}{\mu^{(u,d)}} \right)^2 \; , \nonumber \\ m_{c,s} & = & \frac{\mu^{(u,d)}}{29} \frac{4}{9}\left(80 + \varepsilon^{(u,d)}\right) + \left( A^{(u,d)} - B^{(u,d)} \right)\left( \frac{\alpha^{(u,d)}}{\mu^{(u,d)} }\right)^2 \; , \nonumber \\ m_{t,b} & = & \frac{\mu^{(u,d)}}{29} \frac{24}{25} \left(624 + \varepsilon^{(u,d)} + C^{(u,d)} \right) + B^{(u,d)}\left( \frac{\alpha^{(u,d)}}{\mu^{(u,d)}}\right)^2 \;,
\end{eqnarray}

\ni where

\begin{equation}
A^{(u,d)} = \frac{\mu^{(u,d)}}{29}\,\frac{36}{320 - 5\varepsilon^{(u,d)} }\;
\;,\;\; B^{(u,d)} = \frac{\mu^{(u,d)}}{29}\,\frac{10800}{31696 + 54 C^{(u,d)}
+ 29\varepsilon^{(u,d)}}\;.
\end{equation}

\ni In Eqs. (6), the relative smallness of perturbating terms is more pronounced due to extra factors [{\it cf.} Eqs. (35) further on]. In our discussion, we will take for experimental quark masses the arithmetic means of their lower and upper limits quoted in the Review of Particle Physics [3] {\it i.e.},

\begin{equation}
m_u = 3 \,{\rm MeV}\;,\; m_c = 1.25 \,{\rm GeV}\;,\;m_t = 174 \,{\rm GeV}
\end{equation}

\ni and

\vspace{-0.2cm}

\begin{equation}
m_d = 6 \,{\rm MeV}\;,\; m_s = 123 \,{\rm MeV}\;,\;m_b = 4.2 \,{\rm GeV}\;.
\end{equation}

 Eliminating from the unperturbed terms in Eqs. (6) the constants $\mu^{(u,d)} $ and $\varepsilon^{(u,d)}$, we derive the correlating formulae being counterparts of Eqs. (2) for charged leptons:

\begin{eqnarray}
m_{t,b} & = & \frac{6}{125} \left( 351 m_{c,s} - 136 m_{u,d} \right) + \frac{\mu^{(u,d)}}{29} \frac{24}{25} C^{(u,d)} \nonumber \\ & & - \frac{1}{125} \left(2922 A^{(u,d)} - 2231 B^{(u,d)}\right) \left(\frac{\alpha^{(u,d)}}{\mu^{(u,d)}}\right)^2\;, \nonumber \\ \mu^{(u,d)} & = & \frac{29}{320} \left(9 m_{c,s} - 4m_{u,d}\right) - \frac{29}{320} \left(5 A^{(u,d)} - 9 B^{(u,d)}\right) \left(\frac{\alpha^{(u,d
)}}{\mu^{(u,d)}}\right)^2 \;, \nonumber \\  \varepsilon^{(u,d)} & = & \frac{29 m_{u,d}}{\mu^{(u,d)}} + \frac{29}{\mu^{(u,d)}} A^{(u,d)} \left(\frac{\alpha^{(u,d)}}{\mu^{(u,d)}}\right)^2 \;. 
\end{eqnarray}

\ni The unperturbed parts of these relations are:

\begin{eqnarray}
\stackrel{\circ}{m}_{t,b} & = & \frac{6}{125} \left( 351 m_{c,s} - 136 m_{u,d} \right) + \frac{\stackrel{\circ}{\mu}^{(u,d)}}{29} \frac{24}{25} \stackrel{\circ}{C}^{(u,d)} \nonumber \\ & = & \left\{\begin{array}{c} 21.0 \\ 2.03 \end{array}\right\}\,{\rm GeV} + \frac{\stackrel{\circ}{\mu}^{(u,
d)}}{29}\frac{24}{25} \stackrel{\circ}{C}^{(u,d)} \;, \nonumber \\ \stackrel{\circ}{\mu}^{(u,d)} & = & \frac{29}{320} \left(9 m_{c,s} - 4m_{u,d} \right) = \left\{\begin{array}{c} 1020 \\ 98.1 \end{array} \right\} \,{\rm MeV} \;, \nonumber \\  
\stackrel{\circ}{\varepsilon}^{(u,d)} & = & \frac{29 m_{u,d}}{\stackrel{ \circ}{\mu}^{(u,d)}} = \left\{\begin{array}{l} 0.0854 \\ 1.77 \end{array}
\right\} \;.
\end{eqnarray}

\ni In the spirit of our perturbative approach, the "coupling" constant $ \alpha^{(u,d)}$ can be put zero in all perturbing terms in Eqs. (6) and (10), except for $\alpha^{(u,d)\,2}$ in the numerator of the factor $(\alpha^{(u,d)} /\mu^{(u,d)})^2$ that now becomes $ (\alpha^{(u,d)} / \stackrel{\circ}{\mu}^{(u,d)})^2$. Then, $ A^{(u,d)}$ and $ B^{(u,d)}$ are replaced by 

\begin{equation}
\stackrel{\circ}{A}^{(u,d)} = \frac{\stackrel{\circ}{\mu}^{(u,d)}}{29} \frac{36}{320 - 5\stackrel{\circ}{\varepsilon}^{(u,d)} }\; \;,\;\; \stackrel{\circ}{B}^{(u,d)} = \frac{\stackrel{\circ}{\mu}^{(u,d)}}{29} \frac{10800}{31696 + 54 \stackrel{\circ}{C}^{(u,d)} + 29\stackrel{\circ}{\varepsilon}^{(u,d)}}\;.
\end{equation}

\ni Note that the first Eq. (6) can be rewritten identically as $m_{u,d} =\;\stackrel{\circ}{\mu}^{(u,d)} \,\stackrel{\circ}{\varepsilon}^{(u,d)}\!\!\!/29 $ according to the third Eq. (11). 

We shall be able to return to the discussion of quark masses after an estimation of constants $\alpha^{(u)}$ and $\alpha^{(d)}$ is made. Then, we shall determine the parameters $ C^{(u)}$ and $C^{(d)}$ (as well as their unperturbed parts ${\stackrel{\circ}{C}}^{(u)}$ and ${\stackrel{\circ}{C} }^{(d)} $) playing here an essential role in providing large values for $ m_t $ and $ m_b $.

\vspace{0.3cm}

\ni {\centerline{\bf 3. \CKM matrix}}

\vspace{0.3cm}

At present, we find the unitary matrices $\left(U_{\alpha \beta}^{(u,d)}\right)$ that diagonalize the mass matrices $\left(M_{\alpha \beta}^{(u,d)}\right)$ according to the relations $ U^{(u,d) \,\dagger} M^{(u,d)}U^{(u,d)} = $ diag$(m_{u,d}\,,\,m_{c,s}\,,\,m_{t,b})$. In the lowest perturbative order, the result has the form (4) with the necessary replacement of labels:

\begin{equation}
(e) \rightarrow (u)\;\;{\rm or}\;\;(d)\;,\;\mu \rightarrow c\;\;{\rm or}\;\;s
\;,\;\tau \rightarrow t\;\;{\rm or}\;\;b\;,
\end{equation}

\ni respectively.

 Then, the elements $ V_{\alpha \beta}$ of the \CKM matrix $ V = U^{(u)\,\dagger}U^{(d)}$ 
can be calculated with the use of Eqs. (13) in the lowest perturbative order. Six resulting off--diagonal elements are:

\begin{eqnarray}
V_{us} & = & -V^*_{cd} = \frac{2}{29}\left(\frac{\alpha^{(d)}}{m_s} e^{i\varphi^{(d)}} - \frac{\alpha^{(u)}}{m_c} e^{i\varphi^{(u)}} \right) \;, \nonumber \\ 
V_{cb} & = & -V^*_{ts} = \frac{8\sqrt{3}}{29}\left(\frac{\alpha^{(d)}}{m_b} e^{i\varphi^{(d)}} - \frac{\alpha^{(u)}}{m_t} e^{i\varphi^{(u)}} \right) \simeq \frac{8\sqrt{3}}{29} \frac{\alpha^{(d)}}{m_b} e^{i\varphi^{(d)}} \;, \nonumber \\
V_{ub} & \simeq & -\frac{16\sqrt{3}}{841}\frac{\alpha^{(u)}\alpha^{(d)} }{m_c m_b} e^{i(\varphi^{(u)}+\varphi^{(d)})} \;, \nonumber \\ 
V_{td} & \simeq & \frac{16\sqrt{3}}{841} \frac{\alpha^{(d)\,2}}{m_s m_b}\,e^{-2i\varphi^{(d)}} \;,
\end{eqnarray}

\ni where the indicated approximate steps were made due to the inequality $ m_t \gg m_b $ and/or under the assumption that $\alpha^{(u)}/m_c \gg \alpha^{(d)} /m_b $ ({\it cf.} the conjecture (18) later on). All three diagonal elements are real and positive in a good approximation:

\begin{equation}
V_{ud} \simeq 1 - \frac{1}{2}|V_{us}|^2\;,\;V_{cs} \simeq 1 - \frac{1}{2} |V_{us}|^2 - \frac{1}{2}|V_{cb}|^2\;,\;V_{tb} \simeq 1 - \frac{1}{2}|V_{cb}|^2 \;.
\end{equation}

\ni In fact, in the lowest perturbative order,

\begin{equation}
\arg V_{ud} \simeq \frac{4}{841} \frac{\alpha^{(u)}\alpha^{(d)}}{m_c m_s} \sin \left(\varphi^{(u)} - \varphi^{(d)}\right)\frac{180^\circ}{\pi} \simeq -\arg V_{cs}\;,\;\arg V_{tb} \simeq 0 \;,
\end{equation}

\ni what gives a nearly vanishing $ \arg V_{ud} = 0.88^\circ = -\arg V_{cs}$, if the values (17), 
(19) and (22) are used.

 Taking as an input the experimental value $|V_{cb}| = 0.0402 \pm 0.0019 $ [3], we estimate from the second Eq. (14) that 

\vspace{-0.2cm}

\begin{equation}
\alpha^{(d)} \simeq \frac{29}{8\sqrt{3}}\, m_b\, |V_{cb}| = (353 \pm 17)\;{\rm 
MeV} \;,
\end{equation}

\ni where $ m_b = 4.2 $ GeV. In order to estimate also $\alpha^{(u)}$, we will
tentatively conjecture the approximate proportion

\vspace{-0.2cm}

\begin{equation}
\alpha^{(u)} : \alpha^{(d)} \simeq Q^{(u)\,2} : Q^{(d)\,2} = 4
\end{equation}

\ni to hold, where $ Q^{(u)} = 2/3 $ and $ Q^{(d)} = -1/3 $ are quark electric charges.  Under the conjecture (18)

\vspace{-0.2cm}

\begin{equation}
\alpha^{(u)} \simeq (1410 \pm 70)\, {\rm MeV} \;.
\end{equation}

\ni In this case, from the second and third Eq. (14) we obtain the prediction

\begin{equation}
|V_{ub}|/|V_{cb}| \simeq \frac{2}{29}\frac{\alpha^{(u)}}{m_c} \simeq 0.0779 \pm 0.0037 \;,
\end{equation}

\ni where $ m_c = 1.25 $ GeV. This is consistent with the experimental figure $|V_{ub}|/|V_{cb}| = 0.08 \pm 0.02 $ as well as $ 0.090 \pm 0.025$ [3].

Now, with the experimental value $|V_{us}| = 0.2196 \pm 0.0023$ [3] as another input, we can calculate from the first Eq. (14) the phase difference $\varphi^{(u)} - \varphi^{(d)}$. In fact, taking the absolute value of this equation, we get

\begin{equation}
\cos\left(\varphi^{(u)} - \varphi^{(d)}\right) = \frac{1}{8}\frac{m_c}{m_s} \left[1 + 16 \left( \frac{m_s}{m_c} \right)^2 - \frac{841}{4} \left( \frac{m_s}{\alpha^{(d)}}\right)^2 |V_{us}|^2 \right] = - 0.0967
\end{equation}

\ni with $ m_c = 1.25 $ GeV  and $ m_s = 123 $ MeV, if the proportion (18) is taken into account. Here, the central values of $\alpha^{(d)}$ and $|V_{us}|$ were used. Hence,

\begin{equation}
\varphi^{(u)} - \varphi^{(d)} = 95.5^\circ = -84.5^\circ + 180^\circ \;.
\end{equation}

\ni Then, calculating the argument of the first Eq. (14), we infer that

\begin{equation}
\tan\left(\arg V_{us} - \varphi^{(d)}\right) = -4 \,\frac{m_s}{m_c}\, \frac{\sin\left(\varphi^{(u)} - \varphi^{(d)}\right)}{1 - 4 ({m_s}/{m_c}) \cos\left(\varphi^{(u)} - \varphi^{(d)}\right)} = - 0.377 \;,
\end{equation}

\ni what gives

\vspace{-0.2cm}

\begin{equation}
\arg V_{us} = -20.7^\circ + \varphi^{(d)} \;.
\end{equation}

 The results (22) and (24) together with the formula (14) enable us to evaluate the rephasing--invariant CP--violating phases

\begin{equation}
\arg (V_{us}^*V_{cb}^*V_{ub}) = 20.7^\circ - 84.5^\circ = -63.8^\circ 
\end{equation}

\ni and 

\vspace{-0.2cm}

\begin{equation}
\arg (V_{cd}^*V_{ts}^*V_{td}) = -20.7^\circ \;
\end{equation}

\ni (they are invariant under quark rephasing the same for up and down quarks of the same generation). Note that the sum of arguments (25) and (26) is always equal to $\varphi^{(u)} - \varphi^{(d)} - 180^\circ $. Carrying out quark rephasing (the same for up and down quarks of the same generation), where

\begin{equation}
\arg V_{us} \rightarrow 0 \;,\; \arg V_{cb} \rightarrow 0 \;,\; \arg V_{cd} \rightarrow 180^\circ \;,\; \arg V_{ts} \rightarrow 180^\circ
\end{equation}

\ni and $\arg V_{ud}$, $\arg V_{cs}$, $\arg V_{tb}$ remain unchanged, we conclude from Eqs. (25) and (26) that

\begin{equation}
\arg V_{ub} \rightarrow -63.8^\circ \;,\; \arg V_{td} \rightarrow -20.7^\circ \;.
\end{equation}

\ni The sum of arguments (28) after rephasing (27) is always equal to $\varphi^{(u)} - \varphi^{(d)} - 180^\circ $.

Thus, in this quark phasing, we predict the following \CKM matrix:

\begin{equation}
\left( V_{\alpha \beta} \right) = \left(\begin{array}{ccc}
0.976 & 0.220 & 0.00313\,e^{-i\,63.88^\circ} \\ -0.220 & 0.975 & 0.0402 \\ 0.00795\,e^{-i\,20.7^\circ} & -0.0402 & 0.999 \end{array}\right)\;.
\end{equation}

\ni Here, only $|V_{us}|$ and $|V_{cb}|$ [and quark masses $m_s\,,\;m_c\,,\;m_b $ consistent with the mass matrices $\left( M^{(u)}_{\alpha \beta}\right)$ and $\left( M^{(d)}_{\alpha \beta}\right)$] are our inputs, while all other matrix elements $ V_{\alpha \beta}$, partly induced by unitarity, are evaluated from the relations derived in this Section from the Hermitian mass matrices $\left( M^{(u)}_{\alpha \beta}\right)$ and $\left( M^{(d)}_{\alpha \beta}\right)$ [and the conjectured proportion (18)]. The independent predictions are two, {\it e.g.} for $|V_{ub}|$ and arg$ V_{ub}$, since ten independent quantities (six quark masses, three mixing angles and one CP-violating phase) are parametrized by eight independent model constants ($\mu^{(u)}$, $\mu^{(d)}$, $ \varepsilon^{(u)} $, $\varepsilon^{(d)}$, $ \alpha^{(u)}$ or $\alpha^{(d)}$, $\varphi^{(u)} - \varphi^{(d)}$ and $ C^{(u)}$, $ C^{(d)}$). In Eq. (29), the small phases arising from Eqs. (16), $\arg V_{ud} = 0.9^\circ$ and $\arg V_{cs} = -0.9^\circ$, are neglected (here, arg $(V_{ud}V_{cs}V_{tb}) = 0 $).

The above prediction of $ V_{\alpha \beta}$ implies the following values of Wolfenstein parameters [3]:

\begin{equation}
\lambda = 0.2196\;\;,\;\;A = 0.834 \;\;,\;\;\rho = 0.157\;\;,\;\;\eta = 0.318 
\end{equation}

\ni and of unitary--triangle angles:

\begin{equation}
\gamma = \arctan \frac{\eta}{\rho} = - \arg V_{ub} = 63.8^\circ\;\;,\;\;\beta = \arctan \frac{\eta}{1-\rho} = - \arg V_{td} = 20.7^\circ \;.
\end{equation}

\ni Hence, the predicted value of $\sin 2\beta = 0.661$ is not inconsistent with the recent BaBar experimental result $\sin 2\beta^{\rm exp} = 0.59 \pm 0.14$ [5].

Now, we may turn back to quark masses. From the third Eq. (6) we can evaluate

\begin{equation}
C^{(u,d)} = \frac{29}{\mu^{(u,d)}}\,\frac{25}{24}\,m_{t,b} - 624 - \varepsilon^{(u,d)} - \frac{29}{\mu^{(u,d)}} \,\frac{25}{24}\,B^{(u,d)}\left( \frac{\alpha^{(u,d)}}{\mu^{(u,d)}}\right)^2\;,
\end{equation}

\ni what, in the framework of our perturbative approach, gives

\begin{eqnarray}
C^{(u,d)} & = & \stackrel{\circ}{C}^{(u,d)} + \frac{29}{\stackrel{\circ}{\mu}^{(u,d)}}\, \frac{25}{24} \,m_{t,b} \,\frac{29}{320\stackrel{\circ}{\mu}^{(u,d)}} \,\left(5\stackrel{\circ}{A}^{(u,d)} - 9 \stackrel{\circ}{B}^{(u,d)}\right)\, \left( \frac{\alpha^{(u,d)}}{\stackrel{\circ}{\mu}^{(u,d)}} \right)^2 \nonumber \\ & & - \frac{29}{\stackrel{\circ}{\mu}^{(u,d)}}\,\left( \stackrel{\circ}{A}^{(u,d)} + \stackrel{\circ}{B}^{(u,d)}\right)\,\left( \frac{\alpha^{(u,d)}}{\stackrel{\circ}{\mu}^{(u,d)}} \right)^2\;,
\end{eqnarray}

\ni where

\begin{equation}
\stackrel{\circ}{C}^{(u,d)} = \frac{29}{\stackrel{\circ}{\mu}^{(u,d)}}\, \frac{25}{24}\,m_{t,b} - 624 - \stackrel{\circ}{\varepsilon}^{(u,d)} = \left\{\begin{array}{r} 4540 \\ 667 \end{array}\right\} \;.
\end{equation}

\ni With the central values of $\alpha^{(u)}$ and $\alpha^{(d)}$ as estimated in Eqs. (17) and (19) we find from Eqs. (12)

\begin{equation}
\stackrel{\circ}{A}^{(u,d)}\left(\frac{\alpha^{(u,d)}}{\stackrel{\circ}{\mu}^{(u,d)}}\right)^2 =  \left\{ \begin{array}{r} 7.60 \\ 5.07 \end{array}\right\} \,{\rm MeV}\;,\; \stackrel{\circ}{B}^{(u,d)} \left( \frac{\alpha^{(u,d)}} {\stackrel{\circ}{\mu}^{(u,d)}}\right)^2 = \left\{\begin{array}{r} 2.63 \\ 6.98 \end{array} \right\}\,{\rm MeV} \;,
\end{equation}

\ni where

\begin{equation}
\frac{\stackrel{\circ}{\mu}^{(u,d)}}{29}\left(\frac{\alpha^{(u,d)}}{\stackrel{\circ}{\mu}^{(u,d)}} \right)^2 = \left\{\begin{array}{r} 67.5 \\ 43.8 \end{array}\right\}\,{\rm MeV}\;.
\end{equation}

\ni We calculate from Eqs. (33) with the use of values (35) that

\begin{equation}
C^{(u,d)} = \left\{\begin{array}{r} 4540 \\ 619 \end{array}
\right\} \;.
\end{equation}

 Similarly, from the second and third Eq. (10), making use of the values (35), we obtain

\begin{equation}
{\mu}^{(u,d)} =  \left\{\begin{array}{c} 1020 \\ 102 \end{array}\right\} \,{\rm MeV}\; ,\; \varepsilon^{(u,d)} =  \left\{\begin{array}{c} 0.302 \\ 3.27 \end{array}\right\}\;.
\end{equation}

 We can easily check that, with the values (11) for $\stackrel{\circ}{\mu}^{(u,d)}$ and  $ \stackrel{\circ}{\varepsilon}^{(u,d)}$ and the value (34) for $\stackrel{\circ}{C}^{(u,d)}$ determined as above from quark masses, the unperturbed parts of mass formulae (6) reproduce correctly these masses. In fact, we get numerically

\begin{eqnarray}
\stackrel{\circ}{m}_{u,d} & = &  \frac{\stackrel{\circ}{\mu}^{(u,d)}}{29}\, \stackrel{\circ}{\varepsilon }^{(u,d)} = \left\{ \begin{array}{c} 3 \\ 6 \end{array}\right\}\,{\rm MeV} = m_{u,d}\;, \nonumber \\ \stackrel{\circ}{m}_{c,s} & = &  \frac{\stackrel{\circ}{\mu}^{(u,d)}}{29}\, \frac{4}{9} \left( 80 + \stackrel{\circ}{\varepsilon}^{(u,d)}\right) = \left\{\begin{array}{r} 1250 \\ 123 \end{array} \right\} \,{\rm MeV} = m_{c,s}\;, \nonumber \\ \stackrel{\circ}{m}_{t,b} & = &  \frac{\stackrel{\circ}{ \mu}^{(u,d)}}{29} \,\frac{24}{25}\left(624 + \stackrel{\circ}{\varepsilon}^{(u,d)} + \stackrel{\circ}{ C}^{(u,d)} \right) = \left\{ \begin{array}{c} 174 \\ 4.2 \end{array}\right\}\,{\rm GeV} = m_{t,b}\;.
\end{eqnarray}

\ni The same is true for the unperturbed part of the first correlating formula (10). Here, $\delta m_\alpha = m_\alpha - \stackrel{\circ}{m}_\alpha $ is negligible {\it versus} $ m_\alpha = m^{\rm exp}_\alpha $ ($\alpha = u\,,\,c\,,\,t $ and $ d\,,\,s\,,\,b $).

We would like to stress that, in contrast to the case of charged leptons, where (in the zero perturbative order) $m_\tau $ has been predicted from $m_e $ and $m_\mu $, in the case of up and down quarks two extra parameters $ C^{(u)}$ and $ C^{(d)}$ appear necessarily to provide large masses $m_t $ and $m_b $ (much larger than $m_\tau $). They cause that (even in the zero perturbative order) $m_t $ ($m_b $) cannot be predicted from $m_u $ and $m_c $ ($m_d $ and $m_s $), till the new parameters are quantitatively understood.

If the ratio $ C^{(u)}/ C^{(d)}$ is equal to $\!x\! $, we can write $ C^{(u,d)} = {\rm const}[Q^{(u,d)2} + (3x-4)Q^{(u,d)}B + (3x + 4) B^2]$, where $ Q^{(u,d)} = 2/3\,,\,-1/3 $ and $ B = 1/3$. In the case of Eq. (37) $x = 7.33 = 22/3$. Thus, the baryon number $B$ may be interpreted as contributing largely to the constants $ C^{(u,d)}$.

\vspace{0.3cm}

\ni {\centerline{\bf 4. A possible phase conjecture}}

\vspace{0.3cm}

Note that a conjecture about $ C^{(u)}$ and $ C^{(d)}$ might lead to a prediction for quark masses and so, introduce changes in the "experimental" quark masses (8) and (9) accepted here. The same is true for a conjecture about $ \varphi^{(u)}$ and $\varphi^{(d)}$.

 For instance, the conjecture that the phase difference $\varphi^{(u)} - \varphi^{(d)}$ is maximal,

\begin{equation}
\varphi^{(u)} - \varphi^{(d)} = 90^\circ \;,
\end{equation}

\ni leads through the first equality in Eq. (21) to the condition

\begin{equation}
1 + 16\left(\frac{m_s}{m_c}\right)^2 - \frac{841}{4}\left(\frac{m_s}{
\alpha^{(d)}}\right)^2 |V_{us}|^2 = 0
\end{equation}

\ni predicting for $ s $ quark the mass

\begin{equation}
m_s = 119\,{\rm MeV} 
\end{equation}

\ni (with $\alpha^{(d)} = 353 $ MeV), being only slightly lower than the value 123 MeV used previously. Here, $ m_c $ and $ m_b $ are kept equal to 1.25 and 4.2 GeV, respectively (also masses of $ u\,,\,d $ and $t $ quarks are not changed, while $\stackrel{\circ}{\mu}^{(d)}$, $\stackrel{\circ}{\varepsilon}^{(d)}$ and $\stackrel{\circ}{C}^{(d)}$ change slightly). Then, from the first equality in Eq. (23)

\begin{equation}
\tan\left(\arg V_{us} - \varphi^{(d)}\right) = -4 \,\frac{m_s}{m_c} = - 0.379 \;\;,\;\;\arg V_{us} = -20.8^\circ + \varphi^{(d)} \;.
\end{equation}

\ni After rephasing (27), this gives $\arg V_{ub} + \arg V_{td} = \varphi^{(u)} - \varphi^{(d)} - 180^\circ = -90^\circ $, where

\begin{equation}
\arg V_{ub} = -69.2^\circ\;\;,\;\;\arg V_{td} = -20.8^\circ
\end{equation}

\ni {\it i.e.}, practically $-70^\circ$ and $-20^\circ$. For the new value (42) of $ m_s$, in the approximation used, all $ |V_{\alpha \beta}|$ remain unchanged (with our inputs of $ |V_{us}| = 0.2196 $ and $ |V_{cb}| = 0.0402 $), except for $ |V_{td}| $ which changes slightly, becoming

\begin{equation}
|V_{td}|  = 0.00849\;.
\end{equation}

\ni Thus, in the \CKM matrix predicted in Eq. (29), only $ |V_{td}| $ and the phases (44) show some changes. The Wolfenstein parameters are

\begin{equation}
\rho = 0.126 \;\;,\;\; \eta = 0.332\;,
\end{equation}

\ni while $\lambda $ and $ A $ do not change (here, the sum $\rho^2 + \eta^2 = 0.126$ is also unchanged). Hence, $\gamma + \beta = 90^\circ $ and $\alpha = 180^\circ - \gamma - \beta = 90^\circ $, where

\begin{equation}
\gamma = \arctan \frac{\eta}{\rho} = - \arg V_{ub} = 69.2^\circ \;\;,\;\;\beta = \arctan \frac{\eta}{1-\rho} = - \arg V_{td} = 20.8^\circ\;.
\end{equation}

\ni So, in the case of conjecture (40), the new restrictive relation

\begin{equation}
\frac{\eta}{\rho} = \frac{1 - \rho}{\eta} \;{\rm or}\; \rho^2 + \eta^2 = \rho
\end{equation}

\ni holds, implying the prediction

\begin{equation}
|V_{td}| /|V_{ub}| = \sqrt{\frac{(1-\rho)^2 + \eta^2}{\rho^2 + \eta^2}} = \frac{\eta}{\rho} = 2.64 \;,
\end{equation}

\ni due to the definition of $\rho $ and $\eta $ from $ V_{ub}$ and $ V_{td}$. It is in agreement with our figures for $ |V_{td}|$ and $ |V_{ub}|$. Then, the new relationship

\begin{equation}
\frac{1}{4}\frac{m_c}{m_s}= \frac{\alpha^{(d)} m_c}{\alpha^{(u)} m_s} = 
\frac{\eta}{\rho} 
\end{equation}

\ni follows for quark masses $m_c $, $m_s $ and Wolfenstein parameters $\rho $, $\eta $, in consequence of Eqs. (14) and the conjectured proportion (18). Both its sides are really equal for our values of $m_c $, $m_s$ and $\rho $, $\eta $.

Thus, summarizing, we cannot predict quark masses without an {\it additional} knowledge or conjecture about the constants $\mu^{(u,d)}$, $\varepsilon^{(u,d)}$, $ C^{(u,d)}$, $\alpha^{(u,d)}$ and $\varphi^{(u,d)}$ (in particular, the conjecture (40) predicting $ m_s $ may be natural). However, we always describe them correctly. If we describe them {\it jointly} with quark mixing parameters, we obtain two independent predictions {\it e.g.} for $|V_{ub}|$ and $\gamma = - \arg V_{ub}$: the whole \CKM matrix is calculated from the inputs of $|V_{us}|$ and $|V_{cb}| $ [and of quark masses $ m_s $, $ m_c $ and $ m_b $ consistent with the mass matrices $ \left( M^{(u)}_{\alpha \beta}\right) $ and $\left( M^{(d)}_{\alpha \beta}\right) $].

Concluding, we can claim that our charged-lepton form of mass matrix works also in a promising way for up and down quarks. But, it turns out that, in the framework of this leptonic form of mass matrix, the heaviest quarks, $ t $ and $ b $, require an additional mechanism in order to produce the bulk of their masses (here, it is represented by the large constants $ C^{(u)}$ and $ C^{(d)}$). Such a mechanism, however, intervenes into the process of quark mixing only through quark masses (practically $m_t$ and $m_b$) and so, it does not modify for quarks the charged-lepton form of mixing mechanism.


\vspace{0.3cm}

{\centerline{\bf Appendix: Motivation for the mass matrix (1)}}

\vspace{0.3cm}

The form of Dirac mass matrix (1) is based on two assumptions: ({\it i}) the conjecture that all kinds of matter's fundamental particles existing in Nature can be deduced from Dirac's square-root procedure $ \sqrt{p^2} = \Gamma \cdot p $, constrained by an intrinsic Pauli principle, and ({\it ii}) a simple ansatz for the Dirac mass matrix, formulated on the ground of the conjecture ({\it i}).

As is easy to observe, Dirac's square-root procedure leads generically to the sequence $N = 1,2,3,\ldots $ of generalized Dirac equations [2,1]

$$
\left\{ \Gamma^{(N)}\cdot\left[p - g A(x)\right] - M^{(N)}\right\} \psi^{(N)}(x) = 0\,, 
\eqno({\rm A}.1)
$$

\ni where for any $N$ the Dirac algebra 

\vspace{-0.1cm}

$$
\left\{ \Gamma^{(N)}_\mu\,,\,\Gamma^{(N)}_\nu \right\} = 2 g_{\mu \nu}
\eqno({\rm A}.2)
$$

\ni is constructed by means of a Clifford algebra,

\vspace{-0.1cm}

$$
\Gamma^{(N)}_\mu \equiv \frac{1}{\sqrt{N}} \sum^N_{i=1}  \gamma^{(N)}_{i \mu}\;\; , \;\;\left\{ \gamma^{(N)}_{i \mu}\,,\,\gamma^{(N)}_{j \nu} \right\} = 2 \delta_{i j} g_{\mu \nu}
\eqno({\rm A}.3)
$$

\ni with $i\,,\,j = 1,2,\ldots ,N$ and $\mu\,,\,\nu = 0,1,2,3$. Here, the term $g \Gamma^{(N)} \cdot A(x)$ symbolizes the Standard Model gauge coupling, involving $\Gamma^{(N)}_5 \equiv i\Gamma^{(N)}_0 \Gamma^{(N)}_1 \Gamma^{(N)}_2 \Gamma^{(N)}_3 $ as well as the color, weak--isospin and hypercharge matrices (this coupling is absent for sterile particles such as sterile neutrinos). The mass $ M^{(N)}$ is independent of $\Gamma^{(N)}_\mu$. In general, the mass $ M^{(N)}$ should be replaced by a mass matrix of elements $ M^{(N,N')}$ which would couple $\psi^{(N)}(x)$ with all appropriate $\psi^{(N')}(x)$, and it might be natural to assume for $N \neq N'$ that $\left[ \gamma^{(N)}_{i \mu}\, , \, \gamma^{(N')}_{j \nu} \right] = 0$ {\it i.e.}, $\left[ \Gamma^{(N)}_\mu\, , \, \Gamma^{(N')}_\nu \right] = 0$.

The Dirac--type equation (A.1) for any $N$ implies that

$$
\psi^{(N)}(x) = \left( \psi_{\alpha_1\alpha_2 \ldots \alpha_N}^{(N)}(x) \right) \,,
\eqno({\rm A}.4)
$$

\ni where each $\alpha_i = 1,2,3,4 $ is the Dirac bispinor index defined in its chiral representation in which the matrices

$$
\gamma^{(N)}_{j 5} \equiv i \gamma^{(N)}_{j 0} \gamma^{(N)}_{j 1} \gamma^{(N)}_{j 2} \gamma^{(N)}_{j 3} \;\; , \;\; \sigma^{(N)}_{j 3} 
\equiv \frac{i}{2} \left[ \gamma^{(N)}_{j 1} \, , \, \gamma^{(N)}_{j 2} \right]
\eqno({\rm A}.5)
$$

\ni are diagonal (note that all matrices (A.5), both with equal and different $j$'s, commute simultaneously). The wave function or field $\psi^{(N)}(x) $ for any $N$ carries also the Standard Model (composite) label, suppressed in our notation. The mass $ M^{(N)}$ gets also such a label. The Standard Model coupling of physical Higgs bosons should be eventually added to Eq. (A.1) for any $N$.

For $ N = 1$ Eq. (A.1) is, of course, the usual Dirac equation, for $ N = 2$ it is known as the Dirac form [6] of the K\"{a}hler equation [7], while for $ N \geq 3$ Eqs. (A.1) give us new Dirac--type equations [2,1]. All of them describe some spin--halfinteger or spin--integer particles for $N$ odd and $N$ even, respectively. The nature of these particles is the main subject of the present paper ({\it cf.} also Ref. [2,1]).

The Dirac--type matrices $\Gamma^{(N)}_\mu $ for any $N$ can be embedded into the new Clifford algebra

$$
\left\{ \Gamma^{(N)}_{i \mu}\, ,  \,\Gamma^{(N)}_{j \nu} \right\} = 2\delta_{i j} g_{\mu \nu}
\eqno({\rm A}.6)
$$

\ni [isomorphic with the Clifford algebra introduced for $\gamma^{(N)}_{i \mu}$ in Eq. (A.3)], if $\Gamma^{(N)}_{i \mu}$ are defined by the properly normalized Jacobi linear combinations of $\gamma^{(N)}_{i \mu}$. In fact, they are given as

\begin{eqnarray*}
\Gamma^{(N)}_{1 \mu} & \equiv & \Gamma^{(N)}_\mu \equiv \frac{1}{\sqrt{N}} \sum^N_{i=1}   \gamma^{(N)}_{i \mu}\, , \\ \Gamma^{(N)}_{i \mu} & \equiv & \frac{1} {\sqrt{i(i - 1)}} \left[ \gamma^{(N)}_{1 \mu} + \ldots + \gamma^{(N)}_{(i-1) \mu} - (i - 1) \gamma^{(N)}_{i \mu} \right] 
\end{eqnarray*}

\vspace{-1.80cm}

\begin{flushright}
({\rm A}.7)
\end{flushright}

\vspace{0.2cm}

\ni for $ i = 1$ and $ i = 2,\ldots, N$, respectively. So, $\Gamma^{(N)}_{1}$ and $ \Gamma^{(N)}_{2} \, , \, \ldots,\Gamma^{(N)}_{N}$ represent respectively the  "centre--of--mass" and "relative" Dirac--type matrices. Note that the Dirac--type equation (A.1) for any $N$ does not involve the "relative" Dirac--type matrices $\Gamma^{(N)}_{2} \, , \, \ldots, \Gamma^{(N)}_{N}$, solely including the "centre--of--mass" Dirac--type matrix $\Gamma^{(N)}_{1} \equiv \Gamma^{(N)}$. Since $\Gamma^{(N)}_{i} = \sum^N_{j=1} O_{i j} \gamma^{(N)}_{j}$, where the 
$N\times N$ matrix $ O = \left( O_{i j} \right)$ is orthogonal ($O^T = O^{-1}$), we obtain for the total spin tensor the formula

$$
\sum^N_{i=1}  \sigma^{(N)}_{i \mu \nu} = \sum^N_{i=1}  \Sigma^{(N)}_{i \mu \nu}  \,,
\eqno({\rm A}.8)
$$

\ni where

$$ 
\sigma^{(N)}_{j \mu \nu} \equiv \frac{i}{2} \left[ \gamma^{(N)}_{j \mu} \, , \, \gamma^{(N)}_{j \nu} \right] \;\; ,\;\; \Sigma^{(N)}_{j \mu \nu} \equiv \frac{i}{2} \left[ \Gamma^{(N)}_{j \mu} \, , \, \Gamma^{(N)}_{j \nu} \right]\,.
\eqno({\rm A}.9)
$$

\ni Of course, the spin tensor (A.8) is the generator of Lorentz transformations for $\psi^{(N)}(x)$.

It is convenient for any $N$ to pass from the chiral representations for individual $ \gamma^{(N)}_i$'s to the chiral representations for Jacobi $\Gamma^{(N)}_i$'s in which the matrices

$$
\Gamma^{(N)}_{j 5} \equiv i\Gamma^{(N)}_{j 0} \Gamma^{(N)}_{j 1} \Gamma^{(N)}_{j 2} \Gamma^{(N)}_{j 3} \;\; ,\;\; \Sigma^{(N)}_{j 3} \equiv \frac{i}{2} \left[ \Gamma^{(N)}_{j 1} \, , \, \Gamma^{(N)}_{j 2} \right]
\eqno({\rm A}.10)
$$

\vspace{0.15cm}
\ni are diagonal (they all, both with equal and different $j$'s, commute simultaneously). Note that $\Gamma^{(N)}_{1 5} \equiv \Gamma^{(N)}_5 $ is the Dirac--type chiral matrix as it is involved in the Standard Model gauge coupling in the Dirac--type equation (A.1).

Using the new Jacobi chiral representations, the "centre--of--mass" Dirac-type matrices  $\Gamma^{(N)}_{1 \mu} \equiv \Gamma^{(N)}_\mu$ and $ \Gamma^{(N)}_{1 5} \equiv \Gamma^{(N)}_5$ can be taken in the reduced forms

$$
\Gamma^{(N)}_\mu =  \gamma_\mu \otimes \underbrace{ {\bf 1}\otimes \cdots \otimes {\bf 1}}_{ N-1 \;{\rm times}} \;\; , \;\; \Gamma^{(N)}_5 = \gamma_5  \otimes \underbrace{ {\bf 1}\otimes \cdots \otimes {\bf 1}}_{ N-1 \;{\rm times}} \; , 
\eqno({\rm A}.11)
$$

\ni where $\gamma_\mu$, $ \gamma_5 \equiv i  \gamma_0   \gamma_1 \gamma_2   \gamma_3 $ and {\bf 1} are the usual $4\times 4$ Dirac matrices. For instance, the Jacobi $\Gamma^{(N)}_{i \mu}$'s and $\Gamma^{(N)}_{i 5}$'s for $N = 3$ can be chosen as

$$
\begin{array}{lllllll}
\Gamma^{(3)}_{1 \mu} & = & \gamma_\mu \otimes {\bf 1}\otimes  {\bf 1} & , & 
\Gamma^{(3)}_{1 5} & = & \gamma_5 \otimes {\bf 1}\otimes  {\bf 1} \; , \\ \\
\Gamma^{(3)}_{2 \mu} & = & \gamma_5 \otimes i \gamma_5 \gamma_\mu \otimes  {\bf 1} & , & 
\Gamma^{(3)}_{2 5} & = & {\bf 1} \otimes \gamma_5 \otimes {\bf 1} \; , \\ \\
\Gamma^{(3)}_{3 \mu} & = & \gamma_5 \otimes \gamma_5 \otimes \gamma_\mu & , & \Gamma^{(3)}_{3 5} & = &  {\bf 1}  \otimes {\bf 1}\otimes  \gamma_5 \; . \end{array}
$$

\vspace{-1.15cm}

\begin{flushright}
({\rm A}.12)
\end{flushright}

\vspace{0.2cm}

Then, the Dirac--type equation (A.1) for any $N$ can be rewritten in the reduced form

$$
\left\{ \gamma \cdot \left[p - g A(x)\right] - M^{(N)}\right\}_{\alpha_1\beta_1} \psi^{(N)}_{\beta_1 \alpha_2 \ldots \alpha_N}(x) = 0\;,
\eqno({\rm A}.13)
$$

\ni where $\alpha_1$ and $\alpha_2 \,,\, \ldots\,,\, \alpha_N$ are the "centre--of--mass" and "relative" Dirac bispinor indices, respectively (here, $(\gamma \cdot p)_{\alpha_1\beta_1} = \gamma_{\alpha_1\beta_1} \cdot p$ and $\left( M^{(N)} \right)_{\alpha_1\beta_1} = \delta_{\alpha_1\beta_1} M^{(N)}$, but the chiral coupling $g \gamma \cdot A(x)$ involves within $A(x)$ also the matrix $\gamma_5$ ). Note that in the Dirac--type equation (A.13) for any $N>1$ the "relative" indices $\alpha_2 \,,\, \ldots\,,\, \alpha_N$ are free, but still are subjects of Lorentz transformations (for $\alpha_2$ this was known already in the case of Dirac form [6] of K\"{a}hler equation [7] corresponding to our $N = 2$).

Since in Eq. (A.13) the Standard Model gauge fields interact only with the "centre--of--mass" index $\alpha_1$, this is distinguished from the physically unobserved "relative" indices $\alpha_2 \,,\, \ldots\,,\, \alpha_N$. Thus, it was natural for us to conjecture some time ago that the "relative" bispinor  indices $\alpha_2 \,,\, \ldots\,,\, \alpha_N$ are all undistinguishable physical objects obeying Fermi statistics along with the Pauli principle requiring in turn the full antisymmetry of wave function $\psi_{\alpha_1 \alpha_2 \,,\, \ldots\,,\, \alpha_N}(x)$ with respect to $\alpha_2 \,,\, \ldots\,,\, \alpha_N$ [2]. Hence, only five values of $N$ satisfying the condition $N-1\leq 4$ are allowed, namely $N = 1,3,5$ for $N$ odd and $N = 2,4$ for $N$ even. Then, from the postulate of relativity and the probabilistic interpretation of $\psi^{(N)}(x)$ we were able to infer that three $N$ odd and two $N$ even correspond to states with total spin 1/2 and total spin 0, respectively [2,1].

Thus, the Dirac--type equation (A.1), jointly with the "intrinsic Pauli principle", if considered on a fundamental level, justifies the existence in Nature of {\it three and only three} generations of spin--1/2 fundamental fermions ({\it i.e.}, leptons and quarks) coupled to the \SM gauge bosons. In addition, there should exist {\it two and only two} generations of spin--0 fundamental bosons also coupled to the \SM gauge bosons.

For sterile particles, Eq. (A.13) with any $N$ goes over into the free Dirac--type equation

\vspace{-0.2cm} 

$$
\left(\gamma_{\alpha_1\beta_1} \cdot p - \delta_{\alpha_1\beta_1} M^{(N)}\right) \psi^{(N)}_{\beta_1 \alpha_2 \ldots \alpha_N}(x) = 0
\eqno({\rm A}.14)
$$

\ni (as far as only \SM gauge interactions are considered). Here, no Dirac bispinor index $\alpha_i$ is distinguished by the \SM gauge coupling which is absent in this case. The "centre--of mass" index $\alpha_1$ is not distinguished also by its coupling to the particle's four--momentum, since Eq. (A.14) is physically equivalent to the free Klein--Gordon equation

$$
\left( p^2 - M^{(N)\,2}\right) \psi^{(N)}_{\alpha_1 \alpha_2 \ldots \alpha_N}(x) = 0 \;.
\eqno({\rm A}.15)
$$

\ni Thus, in this case the intrinsic Pauli principle requires that $N \leq 4$, leading to $N = 1,3$ for $N$ odd and $N = 2,4$ for $N$ even. Similarly as before, they correspond to states with total spin 1/2 and total spin 0, respectively [8].

Therefore, there should exist {\it two and only two} spin--1/2 sterile fundamental fermions ({\it i.e.}, two sterile neutrinos $\nu_s$ and $\nu'_s$) and, in addition, {\it two and only two} spin--0 sterile fundamental bosons.

The wave functions or fields of active fermions (leptons and quarks) of three generations and sterile neutrinos of two generations can be presented in terms of $\psi^{(N)}_{\alpha_1 \alpha_2 \ldots \alpha_N}(x)$ as follows

\vspace{-0.2cm} 

\begin{eqnarray*} 
\psi^{(f)}_{\alpha_1}(x) & = & \psi^{(1)}_{\alpha_1}(x) \;, \nonumber \\
\psi^{(f')}_{\alpha_1}(x) & = & \frac{1}{4}\left(C^{-1} \gamma_5 \right)_ {\alpha_2 \alpha_3} \psi^{(3)}_{\alpha_1 \alpha_2 \alpha_3}(x) = \psi^{(3)}_{\alpha_1 1 2}(x) = \psi^{(3)}_{\alpha_1 3 4}(x) \;,\nonumber \\
\psi^{(f'')}_{\alpha_1}(x) & = & \frac{1}{24}\varepsilon_{\alpha_2 \alpha_3 \alpha_4 \alpha_5} \psi^{(5)}_{\alpha_1 \alpha_2 \alpha_3 \alpha_4 \alpha_5}(x) = \psi^{(5)}_{\alpha_1 1 2 3 4}(x) 
\end{eqnarray*}  

\vspace{-1.75cm}

\begin{flushright}
(A.16)
\end{flushright}

\vspace{0.1cm} 

\ni and

\vspace{-0.3cm} 

\begin{eqnarray*} 
\!\!\!\! \psi^{(\nu_s)}_{\alpha_2}(x) & \!=\! & \psi^{(1)}_{\alpha_2}(x) \;, \nonumber \\
\!\!\!\! \psi^{(\nu'_s)}_{\alpha_2}(x) & \!=\! & \frac{1}{6}\left(C^{-1} \gamma_5 
\right)_ {\alpha_2 \alpha_3} \varepsilon_{\alpha_3 \alpha_4 \alpha_5 \alpha_6} 
\psi^{(3)}_{\alpha_4 \alpha_5 \alpha_6}(x) = \left\{\begin{array}{rll} \psi^{(3)}_{ 1 3 4}(x) & {\rm for} & \alpha_2 = 1 \\ - \psi^{(3)}_{2 3 4}(x) & {\rm for} & \alpha_2 = 2 \\ 
\psi^{(3)}_{3 1 2}(x) & {\rm for} & \alpha_2 = 3 \\ - \psi^{(3)}_{4 1 2}(x) & {\rm for} & \alpha_2 = 4  \end{array} \right.\!\! ,
\end{eqnarray*} 

\vspace{-2.3cm}

\begin{flushright}
(A.17)
\end{flushright}

\vspace{0.8cm}

\ni respectively, where $ \psi^{(N)}_{\alpha_1 \alpha_2 \ldots \alpha_N}(x) $ for active fermions [Eq. (A.16)] carries also the \SM (composite) label, suppressed in our notation, and $C$ denotes the usual $4\times 4$ charge--conjugation matrix. We can see that due to the full antisymmetry in $\alpha_i $ indices for $i \geq 2$ these wave functions or fields appear (up to the sign) with the multiplicities 1, 4, 24 and 1, 6 , respectively. Thus, for active fermions and sterile neutrinos there is given the weighting matrix

$$ 
\rho^{(a)\,1/2} = \frac{1}{\sqrt{29}}  \left( \begin{array}{ccc}  1  & 0 & 0 \\ 0 & \sqrt4 & 0  \\ 0 & 0 & \sqrt{24} \end{array} \right) 
\eqno({\rm A}.18)
$$

\ni and

$$ 
\rho^{(s)\,1/2} = \frac{1}{\sqrt{7}}  \left( \begin{array}{cc}  1  & 0  \\ 0 & \sqrt6 \end{array} \right) \; ,
\eqno({\rm A}.19)
$$

\ni respectively. Of course, for both weighting matrices Tr $\rho = 1$.

Concluding this part of Appendix, we would like to say that in our approach to generations of fundamental particles Dirac bispinor indices ("algebraic partons") play the role of building blocks of composite states identified as fundamental particles. Any fundamental particle, {\it active} with respect to the \SM gauge interactions, contains {\it one} "active algebraic parton" (coupled to the \SM gauge bosons) and {\it a number} $N-1$ of "sterile algebraic partons" (decoupled from these bosons). Due to the intrinsic Pauli principle obeyed by "sterile algebraic partons", the number $N$ of all "algebraic partons" within a fundamental particle is restricted by the condition $N-1 \leq 4$, so that only $N = 1,2,3,4,5$ are allowed. It turns out that states with $N = 1,3,5$ carry total spin 1/2 and are identified with three generations of leptons and quarks, while states with $N = 2,4$ get total spin 0 and so far are not identified. Any fundamental particle, {\it sterile} with respect to the \SM gauge interactions, contains only a number $N \leq 4$ of "sterile algebraic partons", thus only  $N = 1,2,3,4$ are allowed. States with $N = 1,3$ correspond to total spin 1/2 and have to be identified as two hypothetic sterile neutrinos, while states with $N = 2,4$ have total spin 0 and are still to be identified.

Our algebraic construction may be interpreted {\it either} as ingeneously algebraic (much like the famous Dirac's algebraic discovery of spin 1/2) {\it or} as the summit of an iceberg of really composite states of $N$ spatial partons with spin 1/2 whose Dirac bispinor indices manifest themselves as our "algebraic partons". In the former algebraic option, we avoid automatically the irksome existence problem of new interactions necessary to bind spatial partons within leptons and quarks of the second and third generations. For the latter spatial option see some remarks in the second Ref. [8].

Eventually, we introduce the following explicit ansatz for the Dirac mass matrix [2,1]

$$ 
M^{(f)} = \rho^{(a) 1/2} h^{(f)} \rho^{(a) 1/2} \, ,
\eqno({\rm A}.20)
$$

\ni where

$$ 
h^{(f)} =  \mu^{(f)} N^2 + (\varepsilon^{(f)} - 1) N^{-2} + \alpha^{(f)}(a e^{i\varphi^{(f)} } + a^\dagger e^{-i\varphi^{(f)} })
\eqno({\rm A}.21)
$$

\ni with $\mu^{(f)} > 0$, $\varepsilon^{(f)} > 0$, $\alpha^{(f)} > 0$ and $0 <\varphi^{(f)} <2 \pi$ being parameters. Here, the matrix

$$ 
N  = \left( \begin{array}{ccc} 1 & 0 & 0 \\ 0 & 3 & 0 \\ 0 & 0 & 5 \end{array}\right) = 1 + 2 n  
\eqno({\rm A}.22)
$$

\ni describes the number of all $\alpha_i $ indices (all "algebraic partons") appearing in three fermion generations, while

$$ 
a = \left( \begin{array}{ccc} 0 & 1 & 0 \\ 0 & 0 & \sqrt2 \\ 0 & 0 & 0 \end{array}\right)  \; ,\;a^\dagger = \left( \begin{array}{ccc} 0 & 0 & 0 \\ 1 & 0 & 0 \\ 0 & \sqrt2 & 0 \end{array}\right)  
\eqno({\rm A}.23)
$$

\ni play the role of "truncated" annihilation and creation matrices for index pairs $\alpha_i \alpha_j $ with $ i,j \geq 2$ (pairs of "sterile algebraic partons"):

$$ 
[a\, , \,n] = a \;,\; [a^\dagger\, , \,n] = -a^\dagger\;,\;n = a^\dagger a = \left( \begin{array}{ccc} 0 & 0 & 0 \\ 0 & 1 & 0 \\ 0 & 0 & 2 \end{array}\right) \, ,  
\eqno({\rm A}.24)
$$

\ni where the "truncation" condition $ a^3 = 0 = a^{\dagger \,3}$ is satisfied. The formulae (A.20) and (A.21) give explicitly Eq. (1).

In the case of quarks, the modification (5) can be described by the additional term

$$ 
\frac{1}{8} C^{(f)} (N - 1)(N - 3) N^{-2}
\eqno({\rm A}.25)
$$

\ni to be introduced into the matrix $h^{(f)}$ ($ f = u\,,\,d$) given in Eq. (A.21).

In the mass matrix (A.20), the first term containing $\mu^{(f)} N^2$ may be intuitively interpreted as an interaction of all $N$ "algebraic partons" treated on equal footing, while the second involving $-\mu^{(f)} (1 - \varepsilon^{(f)}) N^{-2}$, as a subtraction term caused by the fact that there is one "active algebraic parton" distinguished (by its external coupling) among all $N$ "algebraic partons" of which $N-1$, as "sterile", are undistinguishable. This distinguished "algebraic parton" appears, therefore, with the probability $[N!/(N-1)!]^{-1} = N^{-1}$ that, when squared, leads to an additional interaction involving $\mu^{(f)}(1 - \varepsilon^{(f)}) N^{-2}$. The latter interaction should be subtracted from the former in order to obtain for $N = 1$ the small matrix element $M^{(f)}_{11} = \mu^{(f)} \varepsilon^{(f)}/29$. The third term in the mass matrix (A.20) containing $\alpha^{(f)} (a + a^\dagger)$ annihilates and creates pairs of "sterile algebraic partons" and so, is responsible in a natural way for mixing of three fermion generations.

\vfill\eject

~~~~
\vspace{0.6cm}

{\centerline{\bf References}}

\vspace{1.0cm}

{\everypar={\hangindent=0.5truecm}
\parindent=0pt\frenchspacing

{\everypar={\hangindent=0.5truecm}
\parindent=0pt\frenchspacing

~1.~W.~Kr\'{o}likowski, in {\it Spinors, Twistors, Clifford Algebras and Quantum Deformations (Proc. 2nd Max Born Symposium 1992)}, eds. Z.~Oziewicz {\it et al.}, Kluwer Acad. Press, 1993; {\it Acta Phys. Pol.} {\bf B 27}, 2121 (1996).

\vspace{0.15cm}

~2.~W. Kr\'{o}likowski,{\it Acta Phys. Pol.} {\bf B 21}, 871 (1990); {\it Phys. Rev.} {\bf D 45}, 3222 (1992); {\it Acta Phys. Pol.} {\bf B 24}, 1149 (1993). 

\vspace{0.15cm}

~3.{\it ~Review of Particle Physics}, {\it Eur. Phys. J.} {\bf C 15}, 1 (2000). 

\vspace{0.15cm}

~4.~{\it Cf. e.g.} W. Kr\'{o}likowski, {\it Acta Phys. Pol.} {\bf B 32}, 2129 (2001), also hep-ph/0103226.

\vspace{0.15cm}

~5.~B. Aubert {\it et al.} (BaBar collaboration), hep-ex/0107013v1.

\vspace{0.15cm}

~6.~T. Banks, Y. Dothan and D.~Horn, {\it Phys. Lett.} {\bf B 117}, 413 (1982).

\vspace{0.15cm}

~7.~E. K\"{a}hler, {\it Rendiconti di Matematica} {\bf 21}, 425 (1962); see also D.~Ivanenko and L.~Landau, {\it Z. Phys.} {\bf 48}, 341 (1928).

\vspace{0.15cm}

~8.~W. Kr\'{o}likowski, {\it Acta Phys. Pol.} {\bf B 30}, 227 (1999); in {\it Theoretical Physics Fin de Si\`{e}cle (Proc. 12th Max Born Symposium, 1998}), eds. A.~Borowiec  {\it et al.}, Springer Verlag, 2000,  also hep--ph/9808307.

\vfill\eject

\end{document}